\documentclass[aps,prl,reprint,superscriptaddress,nofootinbib,longbibliography,floatfix]{revtex4-2}

\usepackage{amsmath,amssymb,amsfonts,mathtools}
\usepackage{graphicx}
\usepackage[breaklinks=true,colorlinks,citecolor=blue,linkcolor=blue,urlcolor=blue]{hyperref}\usepackage{siunitx}
\usepackage{multirow}
\usepackage{booktabs}
\usepackage{pdfpages}
\usepackage{xr}
\usepackage{pgffor}
\makeatletter
\AtBeginDocument{\let\LS@rot\@undefined}
\makeatother

\def\supplementfilename{supplementary}

\pdfximage{\supplementfilename.pdf}
\def\numbersupplementpages{\the\pdflastximagepages}

\newif\ifarXiv
\arXivtrue 
\usepackage{pdfpages}
\usepackage{xcolor}
\newcommand{\Etarget}{E_{\mathrm{target}}}
\newcommand{\Egs}{E_{\mathrm{GS}}}
\newcommand{\Cclose}{\mathcal{C}}

\newcommand{\TTS}{\mathrm{TTS}}
\newcommand{\Phit}{P_{\mathrm{hit}}}
\newcommand{\Ptarget}{p_{\mathrm{target}}}
\newcommand{\epshit}{\epsilon}

\begin{document}

\title{The Quest for Quantum Advantage in Combinatorial Optimization:\\
End-to-end Benchmarking of Quantum Solvers vs.~Multi-core Classical Solvers}

\author{Pranav Chandarana}
\affiliation{Kipu Quantum GmbH, Greifswalderstrasse 212, 10405 Berlin, Germany}
\affiliation{Department of Physical Chemistry, University of the Basque Country UPV/EHU, Bilbao 48080, Spain}

\author{Alejandro Gomez Cadavid}
\affiliation{Kipu Quantum GmbH, Greifswalderstrasse 212, 10405 Berlin, Germany}
\affiliation{Department of Physical Chemistry, University of the Basque Country UPV/EHU, Bilbao 48080, Spain}

\author{Enrique Solano}
\affiliation{Kipu Quantum GmbH, Greifswalderstrasse 212, 10405 Berlin, Germany}

\author{Thorsten Koch} \email{koch@zib.de}
\affiliation{Zuse Institute Berlin}
\affiliation{Technische Universit{\"a}t Berlin}

\author{Stefan Woerner} \email{wor@zurich.ibm.com}
\affiliation{IBM Quantum, IBM Research Europe -- Zurich}

\author{Narendra N. Hegade}
\email{narendrahegade5@gmail.com}
\affiliation{Kipu Quantum GmbH, Greifswalderstrasse 212, 10405 Berlin, Germany}
\affiliation{IDAL, Electronic Engineering Department, ETSE-UV, University of Valencia, Avgda. Universitat s/n, 46100 Burjassot, Valencia, Spain}

\begin{abstract}
We perform an end-to-end benchmark of a hybrid sequential quantum computing (HSQC) solver for higher-order unconstrained binary optimization (HUBO), executed on IBM Heron r3 quantum processors to evaluate the potential of current quantum hardware for combinatorial optimization with sub-second end-to-end runtimes. All reported runtimes include the complete pipeline--from preprocessing to QPU execution and postprocessing--under strict wall-clock accounting. Across 20 benchmark instances, a single hybrid attempt produces high-quality solutions in less than one second, matching the ground-state energy in 14 cases. At the same runtime, CPU-based solvers, including simulated annealing, memetic tabu search, and EasySolve, do not reach the value obtained by HSQC, whereas an enhanced parallel tempering method and the GPU-accelerated solver ABS3 reach or surpass it. These results show that HSQC, executed on a single QPU, can achieve performance competitive with strong classical solvers running on 128 vCPUs or 8 NVIDIA A100 GPUs, while also providing a reproducible system-level benchmark for tracking progress as quantum hardware and hybrid sequential workflows improve.
\end{abstract}

\maketitle
Many combinatorial optimization problems can be formulated as minimizing a cost function defined over binary variables, or equivalently as finding the ground state of an Ising model defined on spin variables. In many relevant settings, the objective contains interactions involving three or more variables simultaneously~\cite{Stein_2023,11249852,Wang2025,Lucas2014}. Such higher-order interactions arise naturally when constraints are encoded directly in the cost function or when multiway correlations are preserved rather than reduced to quadratic form. These higher-order unconstrained binary optimization (HUBO) problems often generate rugged energy landscapes, particularly on frustrated interaction graphs, and can remain challenging even for strong classical heuristics at moderate problem sizes. 

A central question in quantum optimization is whether current quantum processors can provide practical advantages when the full computational pipeline is taken into account. Although some quantum methods can exhibit theoretical advantages in certain regimes~\cite{sciadv.adj5170,PRXQuantum.5.030348, leng2025quantumhamiltoniandescentnonsmooth, Shaydulin2024, GomezCadavid2025ScalingAdvantageQEMTSLABS}, current hardware is constrained by limited connectivity, finite coherence times, and substantial classical pre- and post-processing costs. All these factors can diminish or erase any potential gain~\cite{Katzgraber2015,Albash2018,Shaydulin2024,Abbas2024}. As a result, fair evaluation requires system-level benchmarks that compare complete hybrid workflows against strong classical baselines under identical end-to-end runtime accounting. 

Existing studies point to an important opportunity for further exploration~\cite{Koch2025QuantumOptimizationBenchmarkingLibrary,stopfer2025quantumportfoliooptimizationextensive,PhysRevApplied.23.014045,Albash2018}. Prior work on digital hardware has examined the performance of quantum algorithms on relatively small or native-connectivity problems~\cite{Pelofske2023,Pelofske2024,Chandarana2025RuntimeQuantumAdvantageDQO,sachdeva2026integratederrorsuppressedpipelinequantum}. Comparisons against diverse classical solvers, especially across both CPU and GPU hardware, remain limited in the gate-model setting. This makes it difficult to assess whether currently accessible quantum devices can already contribute to practically relevant optimization stacks. Here we address this gap using a hybrid sequential quantum computing (HSQC) workflow~\cite{Chandarana2025HybridSequentialQuantumComputing}, which is based on the bias-field digitized counterdiabatic quantum optimization (BF-DCQO)~\cite{Cadavid2024,Romero2024} algorithm. 
The workflow combines a classical warm start leveraging simulated annealing (SA), followed by short digitized quantum evolutions, and a classical post-processing based on memetic tabu search (MTS)~\cite{iskay}. 

All reported runtimes include the SA preprocessing, QPU execution time from \textit{ibm\_boston} obtained from the job's metadata, and MTS postprocessing, as discussed in more detail later. Across two families of HUBO instances, HSQC achieves a fast time-to-solution, substantially reducing worst-case latency and remaining competitive with strong classical methods. The main contribution is therefore not a claim of broad asymptotic separation, but a controlled system-level demonstration that a tightly integrated hybrid workflow can already occupy a competitive operating regime. Additional implementation details, pseudocode, and per-instance benchmark results are provided in the Supplemental Material.
\begin{figure*}
\centering
\includegraphics[width=\linewidth]{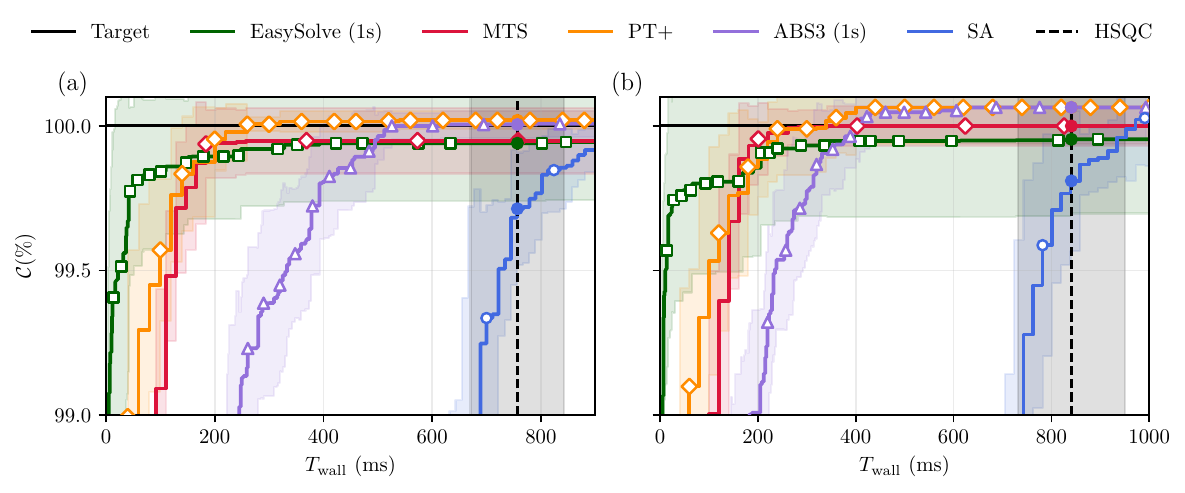}
\caption{Closeness-to-target $\Cclose(t)$ versus wall-clock time for 3S~(left) and 4S~(right).
Curves show means over 10 instances (point-wise best energy over trials/initialization); shaded bands indicate $\pm 1\sigma$.
Vertical dashed line (gray band): HSQC mean runtime $\pm 1\sigma$, $756.9\pm86.2$~ms (3S) and $840.9\pm109.4$~ms (4S).
SA/MTS/EasySolve/PT+: 128 vCPUs, AWS; ABS3: $8\times$~A100~GPUs.
}
\label{fig:closeness_time}
\end{figure*}
\paragraph*{Benchmark design and metrics.}
We study HUBO objectives in Ising form with up to three-local interactions,
\begin{equation}
H(s)=
\sum_{i=1}^{N} J_{i}\, s_i
+
\sum_{(j,k)} J_{jk}\, s_j s_k
+
\sum_{(u,v,w)} J_{uvw}\, s_u s_v s_w ,
\label{eq:hubo_ham}
\end{equation}
where $s_i\in\{+1,-1\}$ and $N$ denotes the number of variables (qubits). The benchmark instances are constructed by swap-layer densification of the heavy-hex coupling graph. At each iteration, parallelizable two- and three-qubit interaction slices are added to a cumulative hypergraph, followed by a SWAP layer that updates the logical-to-physical qubit assignment. We consider two instance families, 3S and 4S, corresponding to $n=3$ and $n=4$ swap layers, respectively, both with $N=156$ variables. The 3S family contains a total of 1128 coupling terms, and the 4S family contains 1323 terms. Coupling strengths are drawn from a standard Cauchy distribution, which produces heavy-tailed frustrated landscapes that are challenging for local search dynamics.

Each HSQC run consists mainly of three stages: classical SA-based warm-start preprocessing, execution of a BF-DCQO circuit on the QPU, and classical MTS-based postprocessing~\cite{Chandarana2025HybridSequentialQuantumComputing}. The exact implementation of the algorithm and the orchestration between the stages of the pipeline are proprietary to Kipu Quantum. The latency of the BF-DCQO stage itself consists of transpilation of the quantum circuit (\(\sim 0.5\,\mathrm{s}\)), submission to Qiskit Runtime (\(\sim 0.8\,\mathrm{s}\)), server-side circuit preparation (\(\sim 3\,\mathrm{s}\) for \texttt{ibm\_boston}), QPU execution (\(\sim 1\,\mathrm{s}\)), and return of the measurement data (\(\sim 2\,\mathrm{s}\)). Since all instances within the considered families share the same transpiled circuit structure and differ only in the rotation angles, transpilation can be performed once offline. Submission and return latencies are deployment dependent.
Server-side circuit preparation is part of the current software stack, however, in principle, it could also be performed offline in a more tightly integrated implementation. 
Accordingly, in the runtime comparisons, we define the quantum runtime contribution to HSQC using the QPU execution interval reported in the IBM Quantum Platform Runtime metadata. End-to-end wall-clock times can then be adjusted using the numbers above to evaluate different scenarios.
The classical AWS baselines are treated analogously: reported runtimes correspond to solver execution only, while SSH or job-launch latency, network transfer, and one-time compilation overhead are excluded.

We benchmark the HSQC pipeline against five classical baselines spanning both CPU and GPU hardware: SA~\cite{kirkpatrick1983}, MTS ~\cite{silva2021quadratic,GALLARDO20091252}, an enhanced in-house implementation of parallel tempering (PT+), and EasySolve~\cite{Nakano2025} were all run on the same AWS bare-metal instance with 64 physical cores (128 vCPUs, 3.9 GHz clock-speed). ABS3~\cite{Nakano2025} was run on 8 NVIDIA A100 GPUs. See the Supplemental Material for more details of the solvers used. For each instance, $\Etarget$ is defined as the minimum energy observed across all HSQC trials and repetition-delay chosen from [20$\mu$s, 40$\mu$s, 80$\mu$s]. Here, the repetition-delay is the waiting time inserted between two consecutive circuit executions (shots) to allow qubits to relax back to the ground state and ensure proper initialization. For all the considered solvers, a run is counted as successful if $E\le \Etarget+\epshit$, with $\epshit=10^{-4}$. The time-to-solution at target success probability $\Ptarget=0.99$ is
\begin{equation}
\TTS=t_{\mathrm{run}}\,\frac{\ln(1-\Ptarget)}{\ln(1-\Phit)},
\label{eq:tts}
\end{equation}
with $\TTS=\infty$ when $\Phit=0$, where $\Phit$ is defined as the probability of a run being successful. For 14 of the 20 instances, $\Etarget$ matches the independently reported ground-state energy $\Egs$ using Gurobi~\cite{gurobi}. For the remaining six, the relative gap $|(\Etarget-\Egs)|/|\Egs|$ is below $0.33\%$. The benchmark, therefore, probes near-ground-state reliability rather than moderate performance. We observe that the main sub-routines in the HSQC pipeline, namely SA, BF-DCQO, and MTS contribute to a monotonic decrease in the average energy across successive stages. These are embedded within a broader workflow that also contains additional lightweight processing steps.

\paragraph*{Sub-second latency regime.}
A practically relevant question is how close competing solvers come to the target energy within the wall-clock time of a single hybrid attempt over ten trials. To address this, we monitor the closeness ratio $\Cclose(t)=E_{\mathrm{best}}(t)/\Etarget$, computed using a best-over-trials/reads aggregation that is optimistic for the classical baselines. At the mean HSQC runtime, $756.9\pm 86.2$~ms in 3S and $840.9\pm 109.4$~ms in 4S, SA remains below $\Cclose=1$ on average, reaching $0.9971$ in 3S and $0.9981$ in 4S. MTS and EasySolve come even closer to the target but still remain marginally below it on average: MTS reaches $0.99949$ in 3S and $0.999996$ in 4S, while EasySolve reaches $0.99940$ in 3S and $0.99954$ in 4S. PT+ and ABS3 are slightly above the target at the same time marker, with PT+ reaching $1.00019$ in 3S and $1.00064$ in 4S, and ABS3 reaching $1.00005$ in 3S and $1.00064$ in 4S. The comparison should be interpreted as a benchmark of complete algorithm--implementation--hardware stacks rather than as a hardware-normalized comparison, especially since ABS3 uses a substantially stronger hardware stack of $8\times$~A100~GPUs. Within the HSQC pipeline, we observe a progressive improvement in $\Cclose(t)$ on average across its stages: from $\sim97\%$ after the SA-based warm start to $\sim98\%$ after the quantum step and $\sim99.4\%$ after MTS-based refinement. Although SA and MTS alone do not reach the target value, their integration with the quantum stage yields improved final solutions.

This short-time gap is not explained by insufficient classical throughput. SA sustains 1.48--1.64~Gbitflips/s and MTS reaches 0.42--3.2~Gbitflips/s in the relevant configurations, indicating that the classical solvers already explore the search space at very high rates. Nevertheless, within the same wall-clock budget, CPU baselines other than PT+ do not reach the hybrid-defined target energy in this near-optimal regime. This identifies a narrow but practically meaningful latency window in which the HSQC pipeline is competitive. The observation is noteworthy because the comparison is made precisely in the regime where practitioners often care most about performance: not after arbitrarily long runs, but under strict latency budgets where near-optimal solutions must be delivered quickly and reliably. In that regime, the hybrid stack benefits not from exhaustive search, but from the coordinated effect of warm starts, shallow quantum evolution, and classical refinement.

\paragraph*{Time-to-solution distributions.}
Table~\ref{tab:tts_summary} summarizes time-to-solution at $\Ptarget=0.99$ for the primary configurations used in comparison: SA with 1000 initialization, MTS with 10 initialization and 5000 generations, and PT+, EasySolve, and ABS3 with a 1s time limit per run. Under these settings, four of the six solvers, namely the HSQC pipeline, SA, PT+, and ABS3, achieve finite TTS on all 20 instances. MTS fails on one instance per family, and EasySolve fails on one 3S instance and four 4S instances. PT+ has the lowest TTS across the benchmark, ranging from 1.00 to 9.02s in both families. We note that standard PT and its variant with isoenergetic cluster moves showed weaker performance and are therefore not reported here. Among the remaining solvers with finite TTS on all 20 instances, the HSQC pipeline shows a markedly tighter range than SA and ABS3 in both families.

For the 3S family, the HSQC pipeline spans $\TTS=3.87$--$43.77$s, compared with 9.80--426.74s for SA, corresponding to nearly an order-of-magnitude reduction in worst-case latency. Instance-by-instance, the HSQC pipeline is faster in 9 of 10 cases, with speedups ranging from $1.34\times$ to $14.26\times$. SA is slightly faster only on a single instance, where the difference is small. For the denser 4S family, HSQC spans 3.78--47.18s, whereas SA spans 10.43--455.13s. Here, the HSQC pipeline is faster in 6 of 10 instances, with speedups ranging from $1.52\times$ to $13.73\times$.

\begin{figure}
\centering
\includegraphics[width=\linewidth]{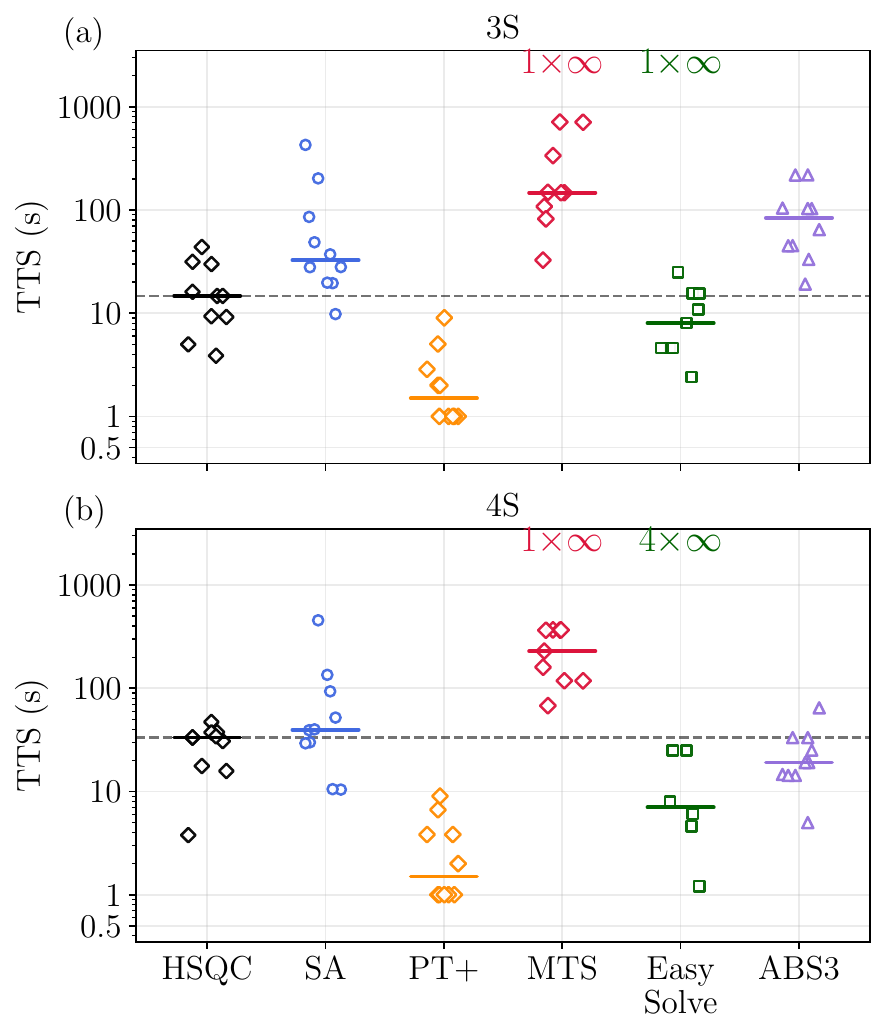}
\caption{Per-instance TTS for 3S~(top) and 4S~(bottom) on a log scale.
Each dot represents one instance; horizontal bars mark medians.
The hybrid solver~(black) occupies a narrow band in both families; SA~(blue) and MTS~(rose) exhibit substantially wider spread.
Failure annotations ($\TTS{=}\infty$) appear above each solver column. The datapoints correspond to the hyperparameters given in Table.~\ref{tab:tts_summary}.}
\label{fig:tts_distribution}
\end{figure}

PT+ remains robust, with finite TTS on all 20 instances and geometric-mean TTS values of 1.87s in 3S and 2.11s in 4S at the 1s setting. In the present TTS data, PT+ is therefore faster than the HSQC pipeline across both families, even though the matched-time closeness comparison of Fig.~\ref{fig:closeness_time} still identifies a sub-second regime in which the hybrid stack reaches near-target quality within a single attempt over 10 trials. The contrast with MTS is stronger: the HSQC pipeline outperforms MTS on all 10 instances in both families, with speedups ranging from $4.66\times$ to $44.14\times$ in 3S and from $1.82\times$ to $97.22\times$ in 4S. Per-instance TTS values on a log scale are reported in Figure~\ref{fig:tts_distribution}.

The geometric-mean TTS, appropriate for multiplicative performance variation, decreases from 45.65s for SA to 13.73s for the HSQC pipeline in 3S, corresponding to a $3.3\times$ speedup. In 4S, the corresponding improvement is from 46.26s to 24.60s, or $1.9\times$. The smaller margin in 4S reflects instances for which SA reaches the target comparatively quickly. PT+ remains lower than the hybrid on this metric, so these results should be read as evidence of a favorable regime relative to SA and MTS rather than proof of universal dominance.
\begin{table}
\centering
\caption{Summary of TTS at $\Ptarget=0.99$ over 10 instances per family. Rows correspond to the primary settings used in the main-text comparison: SA with 1000 initialization, MTS with 10 initialization and 5000 generations, and PT+, EasySolve, and ABS3 with a 1~s runtime limit per run. Reported ranges include only finite values; $\dagger$ marks solvers for which $\TTS=\infty$ occurs on some instances.}
\label{tab:tts_summary}
\setlength{\tabcolsep}{3pt}
\small
\resizebox{\columnwidth}{!}{
\begin{tabular}{llrr}
\toprule
 & Method & $\TTS_{\min}$ (s) & $\TTS_{\max}$ (s) \\
\midrule
\multirow{6}{*}{\rotatebox{90}{3S}}
& HSQC pipeline & 3.87 & 43.77 \\
& SA (1000 initialization)  & 9.80 & 426.74 \\
& PT+ (1s)  & 1.00 & 9.02 \\
& MTS (10, 5000)  & 32.65 & 711.78$^\dagger$ \\
& EasySolve (1s)  & 2.41 & 24.87$^\dagger$ \\
& ABS3 (1s)  & 19.14 & 218.45 \\
\midrule
\multirow{6}{*}{\rotatebox{90}{4S}}
& HSQC pipeline & 3.78 & 47.18 \\
& SA (1000 initialization)  & 10.43 & 455.13 \\
& PT+ (1s)  & 1.00 & 9.02 \\
& MTS (10, 5000)  & 67.78 & 367.30$^\dagger$ \\
& EasySolve (1s)  & 1.21 & 24.88$^\dagger$ \\
& ABS3 (1s)  & 5.00 & 64.52 \\
\bottomrule
\end{tabular}
}
\end{table}
The distributional picture matters as much as the central tendency. EasySolve attains small finite TTS values on several instances, but its failures, one in 3S and four in 4S at the 1s setting, make it less attractive in latency-sensitive settings. ABS3 is competitive on subsets of instances and reaches or exceeds the target in the matched-time comparison, but its TTS range remains broader than the hybrid in 3S and slightly broader in 4S. PT+ is the strongest TTS baseline in the present dataset. The HSQC pipeline, therefore, does not dominate every individual instance or metric, yet it combines reliability with materially smaller TTS than SA and MTS across the full benchmark. From an application perspective, this combination may still matter. In settings where repeated optimization must be executed under fixed service-level objectives, tail latency and failure rates can be as important as average runtime. A solver that occasionally produces the best result but exhibits wide variance or non-negligible failure probability may be less attractive than one that is slightly less aggressive on a subset of instances but consistently reaches the target across the entire workload.
\paragraph*{Sensitivity and interpretation.}
The benchmark is robust to moderate hyperparameter changes, but the details matter. For SA, reducing the number of initializations can introduce failures on several instances, confirming that its reliability depends strongly on repetition count. For PT+, finite TTS becomes common only once the time limit per run reaches the sub-second to one-second regime. For EasySolve and ABS3, extending the time limit per run does not uniformly improve TTS, since success probability does not scale monotonically with runtime on every instance. For the HSQC pipeline, multiple repetition-delay settings were evaluated, and $\Etarget$ is defined as the best result across these settings. 

Two aspects of the benchmark deserve particular care. First, $\Etarget$ is defined using the best energy reached by the HSQC pipeline. This choice does not invalidate the TTS comparison, but it does warrant scrutiny. Its effect is mitigated here by the fact that $\Etarget$ equals the independently reported ground-state energy in most instances and differs only minimally in the rest. Second, the swap-layer construction is deliberately hardware-aware. This makes the benchmark well-suited for testing whether gate-model hardware can be competitive in a native operating regime, but it also limits the extent to which the results should be generalized to unrelated instance ensembles. The paper should therefore be framed as a controlled system-level study rather than as a universal statement about all combinatorial optimization problems.

QAOA is not included because available evidence suggests that low-depth implementations on heavy-hex topologies with higher-order interactions do not outperform SA, while higher-depth variants introduce substantial variational overhead~\cite{Pelofske2023,Pelofske2024}. The non-variational quantum primitive used here avoids this bottleneck and supports a cleaner end-to-end comparison. At the same time, the present benchmark does not isolate the marginal contribution of the quantum stage itself; without an ablation in which the QPU stage is replaced by a classical surrogate, that question remains open. This is an important limitation, but it does not negate the system-level relevance of the present comparison. A full causal decomposition of the performance gain would require additional experiments, including surrogate models for the quantum stage, matched-cost classical replacements, and perhaps instance perturbations to assess how much of the observed benefit is attributable to the hardware-aware structure itself. Those studies fall outside the scope of the present manuscript, but they would sharpen the mechanistic interpretation of the benchmark.

\paragraph*{Discussion and outlook.}
The results do not establish a quantum advantage or an asymptotic separation between quantum and classical optimization. Rather, they show that the HSQC pipeline can be competitive with strong classical solvers under a common solver-side runtime accounting. In this sense, the relevant object of comparison is the overall workflow, including algorithm, implementation, and hardware, rather than the quantum processor in isolation. We also emphasize that this study compares a single QPU against classical solvers executed on multiple CPUs and GPUs. Extending the quantum workflow to multiple QPUs is therefore an important direction for future work. However, the corresponding reduction in time to target is not expected to scale linearly with the number of QPUs, since additional orchestration, communication, and workload-balancing overheads must also be taken into account. A careful analysis of such multi-QPU scaling is left for future work.

Viewed from this workflow-level perspective, the benchmark also highlights the continued strength of highly optimized classical methods. PT+, an enhanced in-house implementation of parallel tempering, achieves the strongest TTS baseline in the present data. EasySolve attains very small finite TTS values on a subset of instances when it succeeds, and ABS3 on $8\times$ A100 GPUs matches or exceeds the target in the matched-time comparison. Highly optimized classical stacks, therefore, remain formidable, and no single solver dominates every metric. The main conclusion is narrower and more defensible: the HSQC pipeline occupies a practically relevant regime, combining broad reliability with substantially reduced tail latency relative to SA and MTS.

More broadly, this work aligns with recent calls for rigorous and practically meaningful benchmarking in quantum optimization. Earlier studies have shown that naively chosen instances may be classically easy~\cite{Katzgraber2015}, that some forms of approximate optimization can display limited advantages in annealing settings~\cite{Albash2018}, and that strong baselines and careful metrics are essential for credible claims~\cite{Abbas2024, Koch2025QuantumOptimizationBenchmarkingLibrary}. The present study contributes to this direction by focusing on higher-order problems, full wall-clock transparency, and comparisons across a diverse set of classical implementations. Just as importantly, it suggests that future discussions of practical quantum advantage may need to move away from narrowly algorithmic notions of speedup and toward a full systems perspective that includes orchestration overhead, target quality, latency tolerance, and implementation realism. Extending the same methodology to larger systems, broader instance classes, and cross-platform comparisons, including annealing-based hardware, is the natural next step.

\textit{Acknowledgements.} The authors thank Sebastián V. Romero for the initial discussions related to this work.
\bibliography{reference}

\ifarXiv
    

\section*{Supplementary material}

# Supplemental Material for "The Quest for Quantum Advantage in Combinatorial Optimization: End-to-end Benchmarking of Quantum Solvers vs. Multi-core Classical Solvers"

Pranav Chandarana,[1, 2] Alejandro Gomez Cadavid,[1, 2] Enrique Solano,[1]
Thorsten Koch,[3, 4, *] Stefan Woerner,[5, †] and Narendra N. Hegade[1, 6, ‡]

[1]*Kipu Quantum GmbH, Greifswalderstrasse 212, 10405 Berlin, Germany*
[2]*Department of Physical Chemistry, University of the Basque Country UPV/EHU, Bilbao 48080, Spain*
[3]*Zuse Institute Berlin*
[4]*Technische Universität Berlin*
[5]*IBM Quantum, IBM Research Europe – Zurich*
[6]*IDAL, Electronic Engineering Department, ETSE-UV, University of Valencia,
Avgda. Universitat s/n, 46100 Burjassot, Spain*

This Supplemental Material provides methodological and implementation details supporting the main Letter. It is organized as follows: (i) problem definitions and the local-energy function used by all solvers, (ii) the hardware-native instance generation strategy (swap-layer densification and coefficient sampling), (iii) baseline solver descriptions and pseudocodes, (iv) metrics and timing methodology, and (v) additional benchmark tables and solver comparisons. We refer to the end-to-end hybrid pipeline as the HSQC. Unless stated otherwise, the main-text summary table uses SA with 1000 initialization, MTS with 10 initializations and 5000 generations. Additionally, 1s per-run timeout for PT+, Easy Solve and ABS3. The additional time-limit settings tabulated below are provided for sensitivity analysis. Finally, $E_{\text{ref}}$ denotes the optimal reference value returned by CPLEX or Gurobi for the corresponding instance.

## I. PROBLEM DEFINITION, REPRESENTATIONS, AND THE LOCAL-ENERGY ORACLE

### A. HUBO objective in binary and Ising form

We consider higher-order unconstrained binary optimization (HUBO). In binary variables $z_i \in \{0, 1\}$, a general HUBO objective of maximum degree $K$ can be written as

$$H_{\text{bin}}(z) = \sum_{\substack{\alpha \subseteq \{1, \dots, N\} \\ 1 \leq |\alpha| \leq K}} c_\alpha \prod_{i \in \alpha} z_i, \tag{S1}$$

where $\alpha$ indexes monomials and $c_\alpha \in \mathbb{R}$ are couplings. All solver implementations used in this work operate internally on an Ising representation $s_i \in \{+1, -1\}$, related to binary variables via $z_i = (1 - s_i)/2$. After this change of variables, the energy becomes a multi-spin Ising polynomial

$$H(s) = \sum_\alpha J_\alpha \prod_{i \in \alpha} s_i + \text{const.}, \tag{S2}$$

with effective couplings $J_\alpha$ that can include terms up to degree $K$. In the experiments summarized in the main text, instance construction generates native two- and three-body interaction support.

### B. Energy evaluation and local flip costs

All classical stages require (i) evaluation of $H(s)$ for a candidate configuration and (ii) rapid evaluation of energy changes for single-spin flips. For a candidate flip of variable $j$ (i.e., $s_j \to -s_j$), only monomials containing $j$ change sign. From Eq. (S2), the exact local energy change is

$$\Delta E_j(s) = H(s^{(j)}) - H(s) = -2s_j \sum_{\alpha: j \in \alpha} J_\alpha \prod_{i \in \alpha \setminus \{j\}} s_i, \tag{S3}$$

---
* koch@zib.de
† wor@zurich.ibm.com
‡ narendrahegade5@gmail.com

where $s^{(j)}$ is the configuration obtained after flipping $j$. We implement an adjacency-list oracle: for each variable $j$, store the list of incident monomials $\{\alpha : j \in \alpha\}$ and update $\Delta E$ using only local neighborhood information. After an accepted flip at index $j$, only $\Delta E_k$ for variables $k$ that share a monomial with $j$ require updates. This yields per-move costs scaling with local hypergraph degree rather than full problem size.

## II. HARDWARE-NATIVE INSTANCE GENERATION

### A. Design goals and high-level structure

Benchmark instances are designed to satisfy two conditions: they are hardware-aware (interaction support realizable on heavy-hex with shallow digitized circuits and minimal embedding overhead), and algorithmically challenging (rugged and frustrated landscapes). We generate instances in two steps: (i) construct interaction support (which variables participate in which two- or three-body terms) using swap-layer densification; and (ii) sample coefficients from a heavy-tailed distribution to create broad local energy scales and outlier couplings.

### B. Coupling map and parallelizable interaction sets

Let $C_0$ denote the initial hardware coupling map, modeled as a graph whose vertices are physical qubits and edges represent native two-qubit connectivity. Given a coupling map $C$, we define:

- $P_{2q} = \{P_{2q}^{(1)}, \ldots, P_{2q}^{(M_2)}\}$, where each $P_{2q}^{(m)}$ is a set of pairwise-disjoint edges in $C$ (a matching). Edges in a matching can be executed in parallel.

- $P_{3q} = \{P_{3q}^{(1)}, \ldots, P_{3q}^{(M_3)}\}$, where each $P_{3q}^{(m)}$ is a set of disjoint three-qubit neighborhoods executable in parallel under the device constraints and compilation strategy.

We obtain these collections using a graph-coloring-based scheduling procedure that converts connectivity into a small set of collision-free interaction time slices.

### C. Swap-layer densification

The key mechanism that increases density while remaining hardware-aware is repeated application of a fixed SWAP layer. Each SWAP layer permutes the logical assignment of variables to physical qubits. After permutation, the next round of parallelizable interaction sets is computed and additional two- and three-body terms are added. At each iteration $i = 1, \ldots, n$, we: (i) compute $(P_{2q}, P_{3q})$ for the current map $C$, (ii) select $S_{2q}$ subsets from $P_{2q}$ and add them to $G_{2q}$, (iii) select $S_{3q}$ subsets from $P_{3q}$ and add them to $G_{3q}$, and then (iv) if $i < n$, update the coupling map by swapping along the first matching $P_{2q}^{(1)}$, see Algorithm 1. Families are labeled by the number of swap layers (e.g., 3S and 4S).

---

**Algorithm 1** Instance layout generation (swap-layer densification)

---

**Require:** Number of swap layers $n$; initial coupling map $C_0$; GraphColoring$(C) \mapsto (P_{2q}, P_{3q})$; SwapRegister$(C, I)$.
1: Initialize interaction sets $G_{2q} \leftarrow \emptyset$, $G_{3q} \leftarrow \emptyset$
2: Set coupling map $C \leftarrow C_0$
3: **for** $i = 1$ to $n$ **do**
4:     $(P_{2q}, P_{3q}) \leftarrow$ GraphColoring$(C)$
5:     Select $S_{2q}$ subsets from $P_{2q}$ and add to $G_{2q}$
6:     Select $S_{3q}$ subsets from $P_{3q}$ and add to $G_{3q}$
7:     **if** $i < n$ **then**
8:         $C \leftarrow$ SwapRegister$(C, P_{2q}^{(1)})$
9:     **end if**
10: **end for**
11: **return** $G_{2q}, G_{3q}$

---

    *a. Selection rule.* We use a deterministic rule for reproducibility: at each layer we take the first $S_{2q}$ matchings from $P_{2q}$ and the first $S_{3q}$ triple-sets from $P_{3q}$.

*b. Why the SWAP layer is chosen as $P_{2q}^{(1)}$.* $P_{2q}^{(1)}$ is a matching and therefore a natural depth-1 SWAP layer. Applying SWAP gates along a matching avoids collisions and keeps compilation predictable.

### D. Coefficient sampling: heavy tails for rugged landscapes

After fixing interaction support, we sample couplings from a heavy-tailed distribution to generate occasional large-magnitude outliers. We draw coefficients from a standard Cauchy distribution, $T_{i_1 \cdots i_k} \sim \text{Cauchy}(0, 1)$, with density $f(\xi) = \frac{1}{\pi(1+\xi^2)}$. The slow tail decay frequently produces large couplings, broadening the distribution of local energy changes $\Delta E_j$.

## III. CLASSICAL BASELINE SOLVERS

### A. Memetic tabu search (MTS)

MTS is a population-based metaheuristic that combines evolutionary variation (crossover and mutation) with an aggressive tabu local search used as a refinement operator, see Algorithm 2. In rugged higher-order Ising landscapes, the tabu phase rapidly descends into a locally optimal basin, while the population loop promotes exploration by recombining information across basins. All evaluations use the same Ising polynomial $H(s) = \sum_\alpha J_\alpha \prod_{i \in \alpha} s_i + \text{const.}$ and local flip-cost oracle defined in Sec. I.B.

*a. Local move oracle and fast $\Delta E$.* For a candidate flip at index $j$, the exact energy change is given by Eq. (S3). We implement an adjacency-list oracle: for each variable $j$, store the list of incident monomials $\{\alpha : j \in \alpha\}$. During tabu refinement we additionally maintain per-term products $\prod_{i \in \alpha} s_i$. Flipping $s_j$ changes the sign of every incident term product, so both the energy and the affected neighborhood can be updated in time proportional to the local hypergraph degree rather than the full problem size $N$. This ensures that per-move cost scales with local connectivity.

*b. Tabu refinement.* Given a starting configuration $s_0$, tabu refinement performs up to $T$ iterations of best-admissible single-spin flips. At each iteration we evaluate all $N$ candidate flips via the local oracle and select the admissible flip (i.e., not forbidden by the tabu mechanism) that minimizes the candidate energy. The tabu mechanism prevents short cycles by forbidding recently visited configurations for a bounded tenure. We implement membership checks using a 64-bit hash of the configuration and maintain a bounded first-in–first-out structure with $O(1)$ lookup. The refinement procedure records the best configuration encountered along the trajectory and returns it. If a target threshold $E_{\text{target}}$ is supplied, refinement terminates early once a configuration satisfying $E \le E_{\text{target}} + \epsilon$ is found.

*c. Population loop and variation operators.* MTS maintains a population of size $P$. Initialization is performed either from random bitstrings or from externally supplied warm starts. Each initial individual is tabu-refined before entering the population so that the population represents locally optimized basins. Each generation then produces $P$ offspring: two parents are selected uniformly at random; a one-point crossover forms an offspring; and the offspring undergoes independent bitwise mutation followed by tabu refinement. After offspring insertion, we retain the lowest-energy $P$ individuals (elitist truncation). The process repeats for $G$ generations or until the target criterion is satisfied.

*d. Mutation schedule.* To balance exploration and exploitation, the mutation rate is annealed over generations. Let $r_{\text{start}}$ and $r_{\text{end}}$ denote the initial and final mutation rates. At generation $g = 1, \ldots, G$, we set

$$r_g = r_{\text{end}} + \left(r_{\text{start}} - r_{\text{end}}\right) \frac{\ln(G + 1 - g)}{\ln(G + 1)}, \tag{S4}$$

so that $r_1 \approx r_{\text{start}}$ and $r_G = r_{\text{end}}$. This schedule yields strong mutation in early generations (promoting basin hopping) and reduced mutation near convergence (preserving elite structures).

*e. Parallel multi-starts and timing.* To estimate hit probabilities, MTS is executed across multiple independent random seeds with identical hyperparameters. These runs are embarrassingly parallel and distributed across available CPU threads. Reported wall-clock times include all steps of the MTS pipeline: initialization (including optional warm-start refinement), all evolutionary generations, and all tabu refinement calls. For throughput reporting, we count (i) full neighborhood evaluations performed during tabu iterations and (ii) explicit bit flips due to mutation and accepted local moves, converting these counts to effective bit-flips per second under a single shared monotonic timer.

*f. Fixed configurations used in the main comparisons.* Unless otherwise noted, MTS uses tabu tenure $T = 20$ and population size $P = 50$. For the main-text fixed configuration at initialization $= 10$, we set $G = 5000$ and anneal mutation from $r_{\text{start}} = 0.1$ to $r_{\text{end}} = 0.001$. Additional fixed-budget configurations (e.g., initialization $= 100$

with varying $G$) are reported in Table S1 to quantify sensitivity to evolutionary budget under the same local-search operator.

---

**Algorithm 2** Memetic Tabu Search (MTS)

---

1: **Input:** $P$, $G$, $T$, $(r_{\text{start}}, r_{\text{end}})$, optional $E_{\text{target}}$
2: Initialize population $\mathcal{P}$ (random or warm starts)
3: **for** each individual $s \in \mathcal{P}$ **do**
4: $\quad (s, E) \leftarrow \textsc{TabuRefine}(s, T, E_{\text{target}})$
5: **end for**
6: **for** $g = 1$ to $G$ **do**
7: $\quad$ Set $r_g$ via Eq. (S4)
8: $\quad \mathcal{O} \leftarrow \emptyset$
9: $\quad$ **for** $k = 1$ to $P$ **do**
10: $\quad\quad$ Select parents $p_1, p_2$ uniformly from $\mathcal{P}$
11: $\quad\quad$ One-point crossover to form offspring $c$
12: $\quad\quad$ **for** $i = 1$ to $N$ **do**
13: $\quad\quad\quad$ **if** $u < r_g$ **then** flip $c_i$
14: $\quad\quad\quad$ **end if**
15: $\quad\quad$ **end for**
16: $\quad\quad (c, E_c) \leftarrow \textsc{TabuRefine}(c, T, E_{\text{target}})$
17: $\quad\quad$ Add $(c, E_c)$ to $\mathcal{O}$
18: $\quad\quad$ **if** $E_c \leq E_{\text{target}} + \epsilon$ **then**
19: $\quad\quad\quad$ **break**
20: $\quad\quad$ **end if**
21: $\quad$ **end for**
22: $\quad \mathcal{P} \leftarrow$ best $P$ individuals from $\mathcal{P} \cup \mathcal{O}$
23: $\quad$ **if** any $E \leq E_{\text{target}} + \epsilon$ **then**
24: $\quad\quad$ **break**
25: $\quad$ **end if**
26: **end for**
27: **Output:** lowest-energy individual in $\mathcal{P}$

---

## B. Simulated annealing (SA)

Simulated annealing (SA) is a single-trajectory stochastic local search that performs randomized single-spin Metropolis updates under a decreasing temperature schedule, see Algorithm 3. SA is well suited as a baseline for rugged Ising landscapes because it combines greedy downhill moves with thermally activated uphill moves that enable barrier crossing. In our implementation, SA evaluates the multi-spin Ising polynomial $H(s) = \sum_\alpha J_\alpha \prod_{i \in \alpha} s_i + \text{const.}$ and uses the exact local energy-difference function from Sec. I.B (Eq. (S3)).

*a. Local Metropolis updates and $\Delta E$ with up to 5-body terms.* Within each sweep, spins are visited in randomized order. For a proposed flip at index $i$, we compute the exact energy change $dE = \Delta E_i(s)$ by summing only the terms incident to $i$ (adjacency lists), algebraically equivalent to Eq. (S3) but made explicit for the supported maximum degree. A proposal is accepted if $\Delta E_i < 0$, otherwise it is accepted with probability $\exp(-\beta \, \Delta E_i)$, where $\beta = 1/T$ is the inverse temperature for the current sweep. The energy is then updated incrementally as $E \leftarrow E + \Delta E_i$, and the best-so-far configuration is tracked online.

*b. Cooling schedule and temperature calibration.* For each instance we construct a temperature range from an instance-specific bound on typical local energy scales. Let

$$\Delta E_i^{\max} \equiv 2|h_i| + 2\sum_j |J_{ij}| + 2\sum_{j<k} |J_{ijk}| + 2\sum_{j<k<\ell} |J_{ijk\ell}| + 2\sum_{j<k<\ell<m} |J_{ijk\ell m}|$$

denote a conservative upper bound on the magnitude of single-spin energy changes at site $i$ given the incident couplings. We then set $T_0 = \max_i \Delta E_i^{\max}$ and $T_f = 0.01\, T_0$, and generate an inverse-temperature schedule $\beta_k = 1/T(k)$ over $S$ sweeps. In the main comparisons we use a geometric schedule

$$T(k) = T_0 \left( \frac{T_f}{T_0} \right)^{k/S}, \qquad k = 1, 2, \ldots, S, \tag{S5}$$

and we also support linear and logarithmic variants for diagnostic use.

*c. Warm starts and multi-run parallelism.* SA is executed as $R$ independent runs ("initialization") per instance to estimate the hit probability. Runs can be initialized either from random configurations or from user-provided warm-start bitstrings, where the encoding follows the Ising convention used throughout: bit value "0" corresponds to spin $+1$ and "1" to spin $-1$. When warm starts are provided, the number of runs defaults to the number of supplied initial states. Independent runs are embarrassingly parallel and are distributed across CPU threads. For each run we record the best energy found and its associated bitstring; the aggregate output includes an "energy distribution" list of $\{(\text{bitstring}, E)\}$ pairs across runs, enabling post hoc computation of hit rates and summary statistics.

*d. Timing and throughput accounting.* Reported wall-clock times include initialization, schedule construction, all Metropolis sweeps, and reduction across threads. We additionally record total attempted spin flips (one attempted update per spin per sweep) and report effective throughput as spin-flips per second with respect to both wall-clock time and accumulated CPU time. These counters are used only for throughput reporting and do not affect the Metropolis dynamics.

---

**Algorithm 3** Simulated Annealing (SA) with multi-starts and warm starts

---

1: **Input:** sweeps $S$, runs $R$, schedule $T(k)$, optional warm starts $\{s_0^{(r)}\}$
2: Compute $T_0$ and $T_f$ from instance couplings; build $\beta_k = 1/T(k)$ for $k = 1, \ldots, S$
3: **for** $r = 1$ to $R$ **in parallel do**
4:     Initialize $s \leftarrow s_0^{(r)}$ if provided else random
5:     $E \leftarrow H(s)$; $(s^\star, E^\star) \leftarrow (s, E)$
6:     **for** $k = 1$ to $S$ **do**
7:         $\beta \leftarrow \beta_k$; randomize spin order
8:         **for all** spins $i$ in randomized order **do**
9:             Compute $\Delta E_i(s)$ using the local oracle (Eq. (S3))
10:             **if** $\Delta E_i < 0$ or $u < \exp(-\beta \Delta E_i)$ **then**
11:                 $s_i \leftarrow -s_i$; $E \leftarrow E + \Delta E_i$
12:                 **if** $E < E^\star$ **then** $(s^\star, E^\star) \leftarrow (s, E)$
13:             **end if**
14:             **end if**
15:         **end for**
16:     **end for**
17:     Record $(s^\star, E^\star)$ for run $r$
18: **end for**
19: **Output:** best $(s^\star, E^\star)$ across runs and the list $\{(s_r^\star, E_r^\star)\}_{r=1}^{R}$

---

### C. Easy Solve

EasySolve is treated as an external backend that accepts higher-order polynomial binary objectives and returns incumbent energies over time. We run it under wall-clock time limits of 1 s, 5 s, and 10 s per run and 10 independent trials per instance. We compute TTS using the same $E_{\text{target}}$ thresholds as for all other methods.

### D. ABS3 (QUBO++)

ABS3 results were obtained using the GPU-accelerated ABS3 backend provided by the QUBO++ framework [1], which supports solving QUBO/HUBO objectives and includes ABS3 as its bundled GPU solver. In QUBO++, an `ABS3Solver` instance is constructed for a given HUBO expression; during construction, the expression is converted into an internal representation and loaded into GPU memory, and the number of GPUs to use can be specified explicitly (otherwise all available devices are used by default).

*a. Coefficient scaling and integerization.* To improve numerical robustness and align with the ABS3/QUBO++ integer-coefficient workflow, coefficients are scaled by a fixed factor $S = 10^6$ and converted to integers prior to optimization, then converted back for reporting. Specifically, for each original coefficient $c_\alpha \in \mathbb{R}$ in Eq. (S1), we form

$$\tilde{c}_\alpha = \text{round}(S\, c_\alpha) \in \mathbb{Z}, \qquad S = 10^6, \tag{S6}$$

solve the integerized problem, and report unscaled energies $\tilde{E}/S$ in the original units. This matches the solver-side scaling convention used in our execution environment. The binary-to-Ising convention is kept consistent with the rest of the Supplemental Material: bit value "0" corresponds to spin $+1$ and "1" to spin $-1$.

*b. Execution environment and device selection.* All visible GPUs are made available to ABS3. QUBO++ exposes device-count selection directly at solver construction time, enabling explicit control of multi-GPU usage for a given run. Each optimization is executed under a strict wall-clock limit $\tau \in \{1, 5, 10\}$ s per trial, with 10 independent trials per instance and time limit. The solver is invoked as an external process (command template `qbpp_solve <instance> <time_limit_s>`) and the measured runtime corresponds to the solver process wall time, excluding any post-processing.

*c. Output parsing and energy traces.* For each trial, we parse solver stdout to extract: (i) the reported QUBO++ version string, (ii) the minimum energy and corresponding bitstring at termination, and (iii) optional intermediate progress callbacks that emit $(t, E)$ updates. These intermediate events are converted into a sparse step-function trace $E_{\text{best}}(t)$ indexed in milliseconds: we retain only timestamps at which the best-so-far energy strictly improves (carry-forward semantics between change points). If no intermediate events are emitted, we record the final minimum energy at the trial end time as a degenerate trace. Per-trial records include start/end timestamps in UTC, the solver exit code, wall time, and the best returned configuration.

*d. Trial aggregation and reporting.* For each instance and time limit, we aggregate successful trials and summarize the distribution of lowest energies using best, worst, mean, and standard deviation. Across all instances, we also record global summary counts (successful versus failed trials) and overall energy statistics. The resulting JSON artifact stores full provenance (platform, host, GPU visibility, environment variables passed to the solver) alongside per-trial traces. Time-to-solution (TTS) is then computed from the empirical hit probability $P_{\text{hit}}$ with respect to the common threshold $E_{\text{target}}$ defined in Sec. IV.A, so that ABS3 and all other baselines are compared under identical target definitions and success criteria.

---

**Algorithm 4** ABS3 experiment driver and trace extraction (QUBO++)

---

1: **Input:** instances $\mathcal{I}$, time limits $\mathcal{T}$, trials per block $R$
2: Detect QUBO++ installation; set environment (scaling $S = 10^6$, device selection, output conventions)
3: **for** instance $i \in \mathcal{I}$ **do**
4:     **for** time limit $\tau \in \mathcal{T}$ **do**
5:         **for** trial $r = 1$ to $R$ **do**
6:             Run `qbpp_solve(instance_i, τ)` and measure wall time
7:             Parse stdout for final minimum energy/bitstring and optional $(t, E)$ progress events
8:             Build sparse best-so-far trace by retaining only strict improvements in $E$
9:             Store per-trial record (exit code, wall time, min energy, bitstring, trace)
10:         **end for**
11:         Summarize energies across successful trials (best/worst/mean/stdev) and record best trial
12:     **end for**
13: **end for**
14: **Output:** JSON bundle with per-trial results, traces, provenance, and summaries

---

### E. Enhanced Parallel Tempering (PT+)

The enhanced parallel tempering (PT+) method used in our work extends standard replica exchange Monte Carlo with five local search enhancements that collectively accelerate convergence by 2–4× compared to vanilla PT on the instances studied.

*a. Temperature ladder.* Each PT+ thread maintains $R = 12$ replicas at geometrically spaced temperatures

$$T_r = T_{\text{cold}} \left( \frac{T_{\text{hot}}}{T_{\text{cold}}} \right)^{r/(R-1)}, \qquad r = 0, 1, \ldots, R-1. \tag{S7}$$

The bounds $T_{\text{cold}}$ and $T_{\text{hot}}$ are calibrated automatically for each problem instance. Five random initial configurations are greedily descended to local minima. Around each minimum, uphill single-spin-flip energy differences $\Delta E > 0$ are sampled by proposing 2,000 random flips, accepting each proposal with probability 0.1 to remain within the local basin. Let $\Delta E_{20}$ and $\Delta E_{95}$ denote the 20th and 95th percentiles of the collected positive energy differences. The temperatures are chosen such that

$$\exp\left( -\frac{\Delta E_{20}}{T_{\text{cold}}} \right) \approx 0.05, \qquad \exp\left( -\frac{\Delta E_{95}}{T_{\text{hot}}} \right) \approx 0.60, \tag{S8}$$

and clamped to $T_{\text{cold}} \in [0.05, 1.5]$, $T_{\text{hot}} \in [3\, T_{\text{cold}}, 15.0]$, ensuring a well-separated and numerically stable ladder.

*b. Replica sweeps and exchanges.* At each iteration, every replica performs $S = 5$ Metropolis sweeps at inverse temperature $\beta_r = 1/T_r$, using the local energy-difference oracle. Replica exchanges are then attempted between adjacent temperature slots using an alternating even–odd scheme. For a pair of neighboring slots $(r, r+1)$, the swap is accepted with probability

$$\min\Big(1,\ \exp\big[(\beta_r - \beta_{r+1})(E_r - E_{r+1})\big]\Big), \tag{S9}$$

where $E_r$ denotes the energy of the replica currently occupying slot $r$. Exchanges are implemented via permutation-index swaps rather than copying spin configurations.

*c. Enhancements over vanilla PT.* PT+ modifies five components of the standard algorithm:

1. **Warm-start initialization.** Rather than random spin assignment, each replica is initialized by assigning random spins followed by a greedy descent (up to 200 passes, accepting only strictly energy-lowering flips) to the nearest local minimum. This eliminates the initial transient during which vanilla PT traverses from random configurations to reasonable energies.

2. **Ordered sweeps.** Each Metropolis sweep visits every spin exactly once in a uniformly random order generated by a Fisher–Yates shuffle, rather than sampling $n$ spins independently with replacement. The latter misses approximately $1 - (1 - 1/n)^n \approx 37\%$ of spins per sweep due to birthday-problem collisions, reducing the information gained per sweep.

3. **Global best tracking.** The best-known solution is tracked across all replicas at all temperatures, not only the coldest. Good configurations can transiently appear at higher temperatures during the exchange process before migrating to colder slots.

4. **Periodic greedy polish.** Every 50 exchange rounds, a greedy descent is applied to the coldest replica to rapidly exploit any favorable basin it currently occupies.

5. **Warm restarts on stagnation.** If no improvement to the global best is observed for 150 consecutive exchange rounds, replicas in the hotter half of the ladder ($r \geq R/2$) are reinitialized by copying the current best solution, perturbing ∼15% of spins at random, and applying greedy descent. The best-known configuration is also re-injected into slot $r = R/4$ to reseed exploration. This replaces the fully random reinitialization used in vanilla PT.

*d. Multi-start parallelism and timing.* PT+ launches $N_{\text{threads}} = 64$ fully independent instances in parallel (one per physical core), each with $R = 12$ replicas, for a total of 768 replicas. The main PT loop runs for $0.95\,\tau$ of the wall-clock budget $\tau$; the remaining time is allocated to a greedy descent applied to the best configuration found across all replicas, sweeping in randomized order until no improving flip exists. The aggregate throughput is approximately $2.1 \times 10^9$ spin-flip attempts per second (wall clock) across all 64 threads. All reported wall-clock times are measured using a single shared monotonic timer started prior to temperature calibration, ensuring that calibration, initialization, and the final greedy descent are included in the total runtime. For convergence traces, the best energy across all $N_{\text{threads}}$ instances is recorded at 20 ms intervals.

## IV. ADDITIONAL BENCHMARK TABLES

In this section we present additional benchmarking tables, including the TTS comparison between HSQC and MTS in Table.S1; TTS comparison between HSQC and SA in Table.S2; TTS comparison between HSQC and EasySolve in Table.S3; TTS comparison between HSQC and ABS3 in Table.S4; and TTS comparison between HSQC and PT+ in Table.S5

TABLE S1. Time-to-target solution (TTS) and throughput for 3S and 4S instances comparing HSQC and MTS. Throughput is reported in Gbitflips/s.

| Instance | Swaps | Energies | | HSQC | inits = 10, gens = 5000 | | inits = 100, gens = 100 | | inits = 100, gens = 4000 | |
|---|---|---|---|---|---|---|---|---|---|---|
| | | $E_{ref}$ | $E_{target}$ | TTS [s] | TTS [s] | Gbitflips/s | TTS [s] | Gbitflips/s | TTS [s] | Gbitflips/s |
| 1 | 3S | -177.9298 | -177.8448 | 43.7735 | 707.9364 | 0.482 | 15.6171 | 2.984 | 262.9691 | 3.181 |
| 2 | 3S | -188.0619 | **-188.0619** | 9.1966 | 146.7019 | 0.479 | 15.4917 | 2.868 | 274.5113 | 3.016 |
| 3 | 3S | -183.8041 | -183.6331 | 14.6861 | 335.9418 | 0.478 | 12.2800 | 3.066 | 160.0696 | 3.164 |
| 4 | 3S | -203.8788 | -203.7766 | 9.3297 | 108.0396 | 0.481 | 18.3346 | 3.031 | 177.2715 | 3.111 |
| 5 | 3S | -209.1214 | **-209.1214** | 16.1262 | 711.7764 | 0.476 | 10.2561 | 2.884 | 222.2817 | 3.163 |
| 6 | 3S | -197.2223 | **-197.2223** | 31.4860 | 146.6610 | 0.473 | 23.0807 | 2.851 | 262.2986 | 3.145 |
| 7 | 3S | -196.0317 | **-196.0317** | 4.9888 | 32.6454 | 0.471 | 2.2775 | 3.054 | 67.4119 | 3.159 |
| 8 | 3S | -194.6620 | **-194.6620** | 14.7035 | ∞ | 0.474 | 25.9226 | 2.890 | 3903.0949 | 3.186 |
| 9 | 3S | -214.8797 | **-214.8797** | 29.9286 | 148.2179 | 0.466 | 14.3321 | 2.977 | 263.2541 | 3.157 |
| 10 | 3S | -207.0658 | **-207.0658** | 3.8747 | 82.0015 | 0.469 | 7.0652 | 2.948 | 77.4076 | 3.155 |
| 1 | 4S | -194.2161 | -194.2030 | 17.6551 | 117.9598 | 0.439 | 7.3291 | 2.689 | 125.2693 | 2.960 |
| 2 | 4S | -194.6358 | **-194.6358** | 15.7802 | ∞ | 0.430 | 77.8944 | 2.710 | 2096.8856 | 2.950 |
| 3 | 4S | -193.2699 | **-193.2699** | 37.5606 | 118.0335 | 0.437 | 28.1917 | 2.775 | 409.1140 | 2.927 |
| 4 | 4S | -195.4433 | **-195.4433** | 47.1788 | 367.2963 | 0.419 | 77.1182 | 2.771 | 329.1163 | 2.859 |
| 5 | 4S | -207.4995 | **-207.4995** | 33.1522 | 228.9058 | 0.436 | 15.8758 | 2.678 | 328.7527 | 2.835 |
| 6 | 4S | -189.7140 | -189.1066 | 33.4736 | 365.8977 | 0.432 | 25.0095 | 2.761 | 231.5597 | 2.913 |
| 7 | 4S | -185.0544 | **-185.0544** | 3.7769 | 367.2007 | 0.427 | 13.7138 | 2.698 | 167.8875 | 2.925 |
| 8 | 4S | -211.3489 | **-211.3489** | 30.7366 | 159.8519 | 0.436 | 28.2351 | 2.762 | 383.6030 | 2.842 |
| 9 | 4S | -167.4161 | -166.8948 | 37.2900 | 67.7797 | 0.434 | 5.2691 | 2.610 | 69.2657 | 2.942 |
| 10 | 4S | -196.6677 | **-196.6677** | 34.1865 | 364.9942 | 0.439 | 22.2007 | 2.788 | 430.9322 | 2.920 |

TABLE S2. Time-to-target solution (TTS) and throughput for 3S and 4S instances comparing HSQC and SA. Throughput is reported in Gbitflips/s.

| Instance | Swaps | Energies | | HSQC | SA (1000 initialization ) | | SA (100 initialization ) | |
|---|---|---|---|---|---|---|---|---|
| | | $E_{ref}$ | $E_{target}$ | TTS [s] | TTS [s] | Gbitflips/s | TTS [s] | Gbitflips/s |
| 1 | 3S | -177.9298 | -177.8448 | 43.7735 | 85.6191 | 1.643 | ∞ | 1.333 |
| 2 | 3S | -188.0619 | **-188.0619** | 9.1966 | 37.1449 | 1.607 | ∞ | 1.333 |
| 3 | 3S | -183.8041 | -183.6331 | 14.6861 | 27.8679 | 1.603 | 50.8792 | 1.340 |
| 4 | 3S | -203.8788 | -203.7766 | 9.3297 | 48.6448 | 1.612 | ∞ | 1.341 |
| 5 | 3S | -209.1214 | **-209.1214** | 16.1262 | 202.2557 | 1.592 | ∞ | 1.333 |
| 6 | 3S | -197.2223 | **-197.2223** | 31.4860 | 27.9808 | 1.596 | 51.2291 | 1.331 |
| 7 | 3S | -196.0317 | **-196.0317** | 4.9888 | 19.5152 | 1.599 | 25.4583 | 1.267 |
| 8 | 3S | -194.6620 | **-194.6620** | 14.7035 | 19.7485 | 1.581 | 8.4821 | 1.228 |
| 9 | 3S | -214.8797 | **-214.8797** | 29.9286 | 426.7428 | 1.598 | ∞ | 1.294 |
| 10 | 3S | -207.0658 | **-207.0658** | 3.8747 | 9.8025 | 1.592 | 54.3210 | 1.258 |
| 1 | 4S | -194.2161 | -194.2030 | 17.6551 | 39.1977 | 1.523 | ∞ | 1.255 |
| 2 | 4S | -194.6358 | **-194.6358** | 15.7802 | 93.3489 | 1.507 | 54.0796 | 1.261 |
| 3 | 4S | -193.2699 | **-193.2699** | 37.5606 | 29.9451 | 1.492 | 54.8049 | 1.244 |
| 4 | 4S | -195.4433 | **-195.4433** | 47.1788 | 39.9646 | 1.494 | 54.3377 | 1.255 |
| 5 | 4S | -207.4995 | **-207.4995** | 33.1522 | 455.1332 | 1.499 | 54.1502 | 1.260 |
| 6 | 4S | -189.7140 | -189.1066 | 33.4736 | 10.4268 | 1.497 | 8.6115 | 1.206 |
| 7 | 4S | -185.0544 | **-185.0544** | 3.7769 | 10.5276 | 1.482 | 57.0841 | 1.196 |
| 8 | 4S | -211.3489 | **-211.3489** | 30.7366 | 134.8384 | 1.494 | ∞ | 1.164 |
| 9 | 4S | -167.4161 | -166.8948 | 37.2900 | 29.2830 | 1.525 | 26.6378 | 1.210 |
| 10 | 4S | -196.6677 | **-196.6677** | 34.1865 | 51.9772 | 1.509 | ∞ | 1.143 |

TABLE S3. Time-to-target solution (TTS) comparison between HSQC and Easy Solve for 3S and 4S instances at different per-run time limits.

| Instance | Swaps | Energies | | HSQC | Easy Solve TTS | | |
|---|---|---|---|---|---|---|---|
| | | $E_{\text{ref}}$ | $E_{\text{target}}$ | TTS [s] | limit = 1s | limit = 5s | limit = 10s |
| 1 | 3S | -177.9298 | -177.8448 | 43.7735 | 15.5660 | 15.0355 | 20.4967 |
| 2 | 3S | -188.0619 | **-188.0619** | 9.1966 | 8.0100 | 5.2501 | 10.2481 |
| 3 | 3S | -183.8041 | -183.6331 | 14.6861 | 4.6084 | 5.2472 | 10.2482 |
| 4 | 3S | -203.8788 | -203.7766 | 9.3297 | 2.4094 | 5.2507 | 10.2469 |
| 5 | 3S | -209.1214 | **-209.1214** | 16.1262 | 4.6116 | 5.2446 | 10.2482 |
| 6 | 3S | -197.2223 | **-197.2223** | 31.4860 | 15.5439 | 20.0789 | 20.4971 |
| 7 | 3S | -196.0317 | **-196.0317** | 4.9888 | 4.6065 | 10.4937 | 10.2485 |
| 8 | 3S | -194.6620 | **-194.6620** | 14.7035 | $\infty$ | $\infty$ | $\infty$ |
| 9 | 3S | -214.8797 | **-214.8797** | 29.9286 | 24.8712 | 26.4103 | 51.5056 |
| 10 | 3S | -207.0658 | **-207.0658** | 3.8747 | 10.8615 | 26.3997 | 39.1938 |
| 1 | 4S | -194.2161 | -194.2030 | 17.6551 | 6.0605 | 10.4974 | 10.2498 |
| 2 | 4S | -194.6358 | **-194.6358** | 15.7802 | $\infty$ | 67.7652 | 132.3360 |
| 3 | 4S | -193.2699 | **-193.2699** | 37.5606 | $\infty$ | 34.8652 | 10.2490 |
| 4 | 4S | -195.4433 | **-195.4433** | 47.1788 | $\infty$ | 229.3953 | $\infty$ |
| 5 | 4S | -207.4995 | **-207.4995** | 33.1522 | 24.8827 | 229.3490 | 211.5191 |
| 6 | 4S | -189.7140 | -189.1066 | 33.4736 | 8.0088 | 10.4958 | 20.4987 |
| 7 | 4S | -185.0544 | **-185.0544** | 3.7769 | 4.6056 | 5.2490 | 10.2505 |
| 8 | 4S | -211.3489 | **-211.3489** | 30.7366 | 24.8700 | 108.2984 | 39.2121 |
| 9 | 4S | -167.4161 | -166.8948 | 37.2900 | 1.2058 | 5.2464 | 10.2485 |
| 10 | 4S | -196.6677 | **-196.6677** | 34.1865 | $\infty$ | 229.4116 | 132.3582 |

TABLE S4. Time-to-target solution (TTS) comparison between HSQC and ABS3 fixed for 3S and 4S instances at different per-run time limits.

| Instance | Swaps | Energies | | HSQC | ABS3 TTS | | |
|---|---|---|---|---|---|---|---|
| | | $E_{\text{ref}}$ | $E_{\text{target}}$ | TTS [s] | limit = 1s | limit = 5s | limit = 10s |
| 1 | 3S | -177.9298 | -177.8448 | 43.7735 | 104.3084 | 25.7553 | 180.7673 |
| 2 | 3S | -188.0619 | **-188.0619** | 9.1966 | 64.4495 | 17.9973 | 70.3711 |
| 3 | 3S | -183.8041 | -183.6331 | 14.6861 | 33.2214 | 34.4180 | 70.3388 |
| 4 | 3S | -203.8788 | -203.7766 | 9.3297 | 45.1082 | 59.7622 | 70.4018 |
| 5 | 3S | -209.1214 | **-209.1214** | 16.1262 | 218.4516 | 25.7436 | 40.0637 |
| 6 | 3S | -197.2223 | **-197.2223** | 31.4860 | 103.1747 | 116.1343 | 126.1408 |
| 7 | 3S | -196.0317 | **-196.0317** | 4.9888 | 19.1419 | 45.2362 | 40.0773 |
| 8 | 3S | -194.6620 | **-194.6620** | 14.7035 | 218.1382 | 393.1682 | 611.8714 |
| 9 | 3S | -214.8797 | **-214.8797** | 29.9286 | 103.2546 | 185.7007 | 126.2559 |
| 10 | 3S | -207.0658 | **-207.0658** | 3.8747 | 45.0189 | 45.1945 | 93.0910 |
| 1 | 4S | -194.2161 | -194.2030 | 17.6551 | 14.6117 | 17.9931 | 28.0093 |
| 2 | 4S | -194.6358 | **-194.6358** | 15.7802 | 64.5152 | 59.7430 | $\infty$ |
| 3 | 4S | -193.2699 | **-193.2699** | 37.5606 | 19.0794 | 81.0808 | 40.0646 |
| 4 | 4S | -195.4433 | **-195.4433** | 47.1788 | 33.2463 | 34.4213 | 70.3621 |
| 5 | 4S | -207.4995 | **-207.4995** | 33.1522 | 33.2264 | 17.9764 | 28.0081 |
| 6 | 4S | -189.7140 | -189.1066 | 33.4736 | 25.1598 | 25.7359 | 28.0018 |
| 7 | 4S | -185.0544 | **-185.0544** | 3.7769 | 19.0855 | 9.0092 | 13.9906 |
| 8 | 4S | -211.3489 | **-211.3489** | 30.7366 | 14.2997 | 34.4101 | 27.9985 |
| 9 | 4S | -167.4161 | -166.8948 | 37.2900 | 4.9984 | 9.0012 | 13.9990 |
| 10 | 4S | -196.6677 | **-196.6677** | 34.1865 | 14.3112 | 34.4261 | 40.0500 |

TABLE S5. Time-to-target solution (TTS) comparison between HSQC and PT+ for 3S and 4S instances at different per-run time limits.

| Instance | Swaps | Energies | | HSQC | PT+ TTS | | |
|---|---|---|---|---|---|---|---|
| | | $E_{ref}$ | $E_{target}$ | TTS [s] | limit = 0.5s | limit = 1s | limit = 5s |
| 1 | 3S | -177.9298 | -177.8448 | 43.7735 | 2.5129 | 1.0000 | 5.0000 |
| 2 | 3S | -188.0619 | **-188.0619** | 9.1966 | 3.3219 | 2.8614 | 5.0000 |
| 3 | 3S | -183.8041 | -183.6331 | 14.6861 | 3.3219 | 1.0000 | 5.0000 |
| 4 | 3S | -203.8788 | -203.7766 | 9.3297 | 4.5076 | 2.0000 | 5.0000 |
| 5 | 3S | -209.1214 | **-209.1214** | 16.1262 | 10.3189 | 5.0259 | 5.0000 |
| 6 | 3S | -197.2223 | **-197.2223** | 31.4860 | 1.0000 | 1.0000 | 5.0000 |
| 7 | 3S | -196.0317 | **-196.0317** | 4.9888 | 1.9125 | 1.0000 | 5.0000 |
| 8 | 3S | -194.6620 | **-194.6620** | 14.7035 | 1.4307 | 2.0000 | 5.0000 |
| 9 | 3S | -214.8797 | **-214.8797** | 29.9286 | 21.8543 | 9.0152 | 5.0000 |
| 10 | 3S | -207.0658 | **-207.0658** | 3.8747 | 1.4307 | 1.0000 | 5.0000 |
| 1 | 4S | -194.2161 | -194.2030 | 17.6551 | 21.8543 | 2.0000 | 5.0000 |
| 2 | 4S | -194.6358 | **-194.6358** | 15.7802 | 10.3189 | 3.8250 | 5.0000 |
| 3 | 4S | -193.2699 | **-193.2699** | 37.5606 | 3.3219 | 1.0000 | 5.0000 |
| 4 | 4S | -195.4433 | **-195.4433** | 47.1788 | 10.3189 | 6.6439 | 5.0000 |
| 5 | 4S | -207.4995 | **-207.4995** | 33.1522 | 6.4557 | 1.0000 | 5.0000 |
| 6 | 4S | -189.7140 | -189.1066 | 33.4736 | 0.5000 | 1.0000 | 5.0000 |
| 7 | 4S | -185.0544 | **-185.0544** | 3.7769 | 0.5000 | 1.0000 | 5.0000 |
| 8 | 4S | -211.3489 | **-211.3489** | 30.7366 | 6.4557 | 9.0152 | 5.0000 |
| 9 | 4S | -167.4161 | -166.8948 | 37.2900 | 1.9125 | 1.0000 | 5.0000 |
| 10 | 4S | -196.6677 | **-196.6677** | 34.1865 | $\infty$ | 3.8250 | 5.0000 |

## V. HARDWARE USED FOR BASELINE EXPERIMENTS

TABLE S6. Execution platforms used for the benchmark.

| Component | Specification |
|---|---|
| Gate-model quantum workflow (BF-DCQO / HSQC) | Single QPU execution stream with classical pre- and post-processing on the CPU host; effective sampling rate is approximately 10 kHz in the supplied timing calibrations, depending on repetition delay and shot count. |
| CPU host (SA, MTS, Easy Solve,PT+, hybrid pre/post) | AWS bare-metal instance with 64 physical CPU cores available (128 vC-PUs reported by the system). |
| GPU host (ABS3) | Multi-GPU node with 8 NVIDIA A100 GPUs visible to the solver. |

## VI. AGGREGATION SENSITIVITY AND TIMING CALIBRATION.

The timing metadata supplied with the benchmark show the expected near-linear growth of total QPU time with shot count, with a pronounced dependence on repetition delay, see Fig. S1. These calibration plots support the use of an effective sampling rate of order 10 kHz as an operational description of the present hardware configuration, while also making clear that this rate is a systems quantity rather than a fixed device constant, see also Fig. S1, at a repetition delay of $80\mu s$, the sampling rate of our circuits is 10kHz − 11kHz.

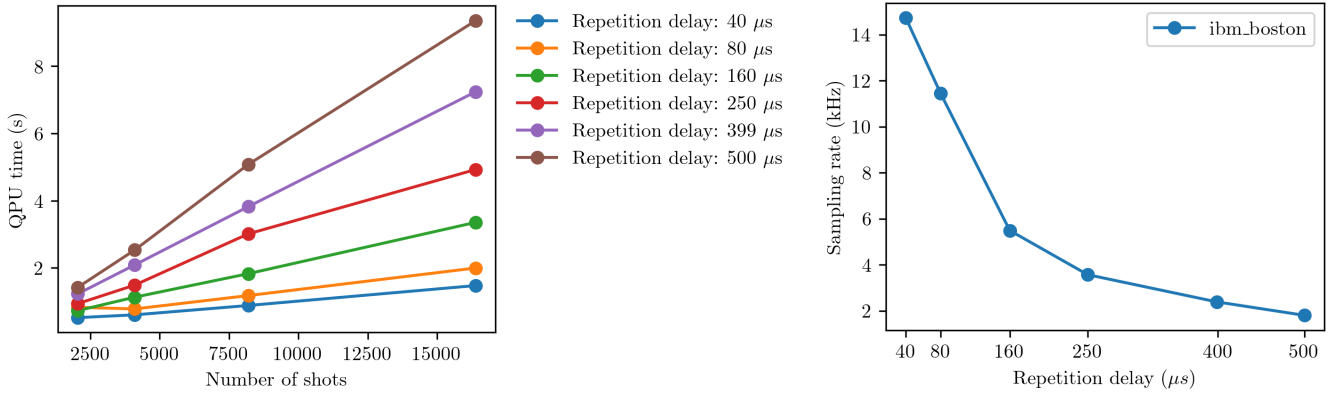

FIG. S1. (Left) QPU time for different number of shots and repetition delay. It grows approximately linearly with shot count for fixed repetition delay, and the slope increases with repetition delay. (Right) QPU sampling rate for different repetition delays. The values are obtained assuming a linear model on the time vs shots plot.


\fi

\end{document}